\documentclass[aps,prd,nofootinbib,floatfix,amsmath,amssymb]{revtex4}
\usepackage{graphicx}
\begin{document}

\makeatletter
\newbox\slashbox \setbox\slashbox=\hbox{$/$}
\newbox\Slashbox \setbox\Slashbox=\hbox{\large$/$}
\def\pFMslash#1{\setbox\@tempboxa=\hbox{$#1$}
  \@tempdima=0.5\wd\slashbox \advance\@tempdima 0.5\wd\@tempboxa
  \copy\slashbox \kern-\@tempdima \box\@tempboxa}
\def\pFMSlash#1{\setbox\@tempboxa=\hbox{$#1$}
  \@tempdima=0.5\wd\Slashbox \advance\@tempdima 0.5\wd\@tempboxa
  \copy\Slashbox \kern-\@tempdima \box\@tempboxa}
\def\FMslash{\protect\pFMslash}
\def\FMSlash{\protect\pFMSlash}
\def\miss#1{\ifmmode{/\mkern-11mu #1}\else{${/\mkern-11mu #1}$}\fi}
\makeatother

\title{Higgs mediated flavor violating top quark decays $t\to u_iH,\, u_i\gamma, \, u_i\gamma \gamma$, and the process $\gamma \gamma \to tc$ in effective theories}

\author{J. I. Aranda$^{(a)}$, A. Cordero-Cid$^{(b)}$, F. Ram\'\i rez-Zavaleta$^{(a)}$, J. J. Toscano$^{(c)}$, E. S. Tututi$^{(a)}$}
\address{$^{(a)}$Facultad de Ciencias F\'\i sico Matem\' aticas,
Universidad Michoacana de San Nicol\' as de
Hidalgo, Avenida Francisco J. M\' ujica S/N, 58060, Morelia, Michoac\'an, M\' exico. \\
$^{(b)}$Facultad de Ciencias de la Electr\' onica, Benem\' erita Universidad
Aut\' onoma de Puebla, Blvd. 18 Sur y Av. San Claudio, 72590, Puebla, Pue., M\' exico.\\
$^{(c)}$Facultad de Ciencias F\'{\i}sico Matem\'aticas,
Benem\'erita Universidad Aut\'onoma de Puebla, Apartado Postal
1152, Puebla, Puebla, M\'exico.}
\begin{abstract}
The rare top quark couplings $tu_i\gamma$ and $tu_i\gamma \gamma$ ($u_i=u,c$) induced at the one-loop level by a flavor violating $tu_iH$ vertex are studied within the context of an effective Yukawa sector that incorporates $SU_L(2)\times U_Y(1)$-invariant operators of up to dimension six. Data on the recently observed $D^0-\overline{D^0}$ mixing are employed to constrain the $tu_iH$ vertex, which is then used to predict the $t\to u_iH$, $t\to u_i\gamma$, and $t\to u_i\gamma \gamma$ decays, as well as the  $\gamma \gamma \to \bar{t}u_i+t\bar{u}_i$ reaction in the context of the ILC. It is found that the $t\to cH$ and $t\to c\gamma \gamma$ decays  can reach sizable branching ratios as high as $5\times 10^{-3}$ and $10^{-4}$, respectively. As for the $t\to c\gamma$ decay, it can have a branching ratio of  $5\times 10^{-8}$ that is about 6 orders of magnitude larger than the standard model prediction, which, however, is still very small to be detected. As for $tc$ production, it is found that, due to the presence of a resonant effect in the convoluted cross section $\sigma(e^+e^-\to \gamma \gamma \to t\bar{c}+\bar{t}c)$, about $(0.5 - 2.7)\times 10^{3}$ $tc$ events may be produced at the ILC for a value of the Higgs mass near to the top mass.
\end{abstract}


\maketitle
The fact that the top quark is the heaviest elementary particle known suggests that it would be quite sensitive to new physics effects. In the standard model (SM), it decays almost exclusively into the $bW$ mode. Although some of the tree-level $t\to d_iW$ ($d_i=d,s$), $t\to d_iWZ$, and $t\to u_iWW$~\cite{EY} ($u_i=u,c$) decays can have sizable branching ratios, they are indeed suppressed. At the one-loop level, there arise the interesting flavor violating decays $t\to u_iV$ ($V=g,\gamma,Z,H$)~\cite{EP,MP}, $t\to cgg$~\cite{EFT}, and $t\to c\bar{c}c$~\cite{CHTT}, which are however considerably GIM-suppressed. Although very suppressed in the SM, the flavor violating $t\to u_iV$ decays can have sizable branching ratios in many well motivated of its extensions, such as the two-Higgs doublet model~\cite{THDM,DPTT}, supersymmetric models~\cite{SUSY-SB,SUSY-RP}, some exotic scenarios~\cite{EM}, 331 models~\cite{CTT}, and model-independent descriptions~\cite{EL1,EL2}. On the other hand, it is expected that new physics effects on top quark physics can be observed in both LHC and ILC colliders, since they complement the capabilities of each other~\cite{ILC, LHC-ILC}. In addition the ILC may be operated in the $\gamma\gamma$ collision mode, which will provide a cleaner environment for top quark physics~\cite{ILC}.

In a recent communication by some of us~\cite{AFRTTT}, exact amplitudes for the one-loop induced $f_if_j\gamma$ and $f_if_j\gamma \gamma$ couplings, with a $f_i$ a quark or charged lepton, induced by a nondiagonal $f_if_jH$ vertex was presented in the context of new sources of flavor violation arising from extended Yukawa sectors. An effective Lagrangian description for the SM Yukawa sector was proposed, which was extended to include dimension-six $SU_L(2)\times U_Y(1)$-invariant operators that generate a general flavor and CP-violating $f_if_jH$ coupling of renormalizable type without the necessity of including additional degrees of freedom. Applications to lepton flavor violation were done in~\cite{AFRTTT}. In this paper, we will apply these general results to study the rare top quark decays $t\to u_iH$, $t\to u_i\gamma$, and $t\to u_i\gamma \gamma$, as well as to investigate the $\gamma \gamma \to \bar{t}u_i+t\bar{u}_i$ reaction in the context of the ILC operating in the $\gamma \gamma$ mode, which has shown to be the most efficient mechanism in producing $tc$ events, as the corresponding cross section can be up to 2 orders of magnitude and 1 order of magnitude greater than those associated with the production mechanism in the $e^+e^-$ and $e\gamma$ collision modes, respectively~\cite{SM,THDM1,THDM2,SUSY1,SUSY2,TCAT,LHM1,LHM2}. In fact, the possibility of detecting $tc$ events at linear $e^+e^-$ colliders operating in the  $e^+e^-$, $e\gamma$, and $\gamma \gamma$ modes has been considered by some authors in diverse contexts. In the SM, there is no signal of $tc$ events as the three collision modes have very small cross sections: $\sigma(e^+e^-\to tc)\approx 10^{-10}\,\mathrm{fb}$, $\sigma(e^-\gamma\to e^-tc)\approx 10^{-9}\,\mathrm{fb}$, and $\sigma(\gamma \gamma \to tc)\approx 10^{-8}\,\mathrm{fb}$~\cite{SM,SUSY2}. In contrast, cross sections several orders of magnitude larger arise within the context of the general two-Higgs doublet model (THDM-III), it was found that $\sigma(e^+e^-\to \nu_e\bar{\nu}_e tc)\approx 9\,\mathrm{fb}$~\cite{THDM1} and  $\sigma(\gamma \gamma \to tc)\approx 0.11\, \mathrm{fb}$~\cite{THDM2}. On the other hand, the cross sections $\sigma(e^+e^-\to tc)\approx 0.02\,\mathrm{fb}$, $\sigma(e^-\gamma\to e^-tc)\approx 0.04\,\mathrm{fb}$, and $\sigma(\gamma \gamma \to tc)\approx 0.7\,\mathrm{fb}$ were reported in the context of supersymmetric models~\cite{SUSY1,SUSY2}. In top-color-assisted technicolor models, the following cross sections were found~\cite{TCAT}: $\sigma(e^+e^-\to \gamma^*\gamma^*,\gamma^* Z^*\to e^+e^- tc)\approx 10^{-1}\,\mathrm{fb}$, $\sigma(e^-\gamma\to e^-tc)\approx 3\,\mathrm{fb}$, and $\sigma(\gamma \gamma \to tc)\approx 50\,\mathrm{fb}$. More promising results have been found in the context of the littlest Higgs model. In Ref.~\cite{LHM1} the cross sections $\sigma(e^+e^-\to tc)\approx 16\,\mathrm{fb}$, $\sigma(e^+e^-\to \nu_e\bar{\nu}_e tc)\approx 10^{-3}\,\mathrm{fb}$ $\sigma(e^-\gamma\to e^-tc)\approx 10^{-2}\,\mathrm{fb}$ were found. More recently, the very large cross sections $\sigma(e^+e^-\to tc)\approx 10^{4}\,\mathrm{fb}$ and  $\sigma(\gamma \gamma \to tc)\approx 10^8\,\mathrm{fb}$ were reported in the context of the littlest Higgs model with $T$-parity~\cite{LHM2}. The capabilities of model-independent $tc$ production at high energy $e^+e^-$ colliders have been studied in Refs.~\cite{MIPee}. Below, we will see that the flavor violating $tcH$ coupling induces an observable cross section for the complete $e^+e^-\to \gamma \gamma \to t\bar{c}+\bar{t}c$ process. On the other hand, experimental analysis on $tc$ events mediated by a vector particle as the photon or the $Z$ gauge boson has been carried out within the context of LEPII and HERA~\cite{VM}. Constraints on the $tc\gamma$ and $tcZ$ couplings of order of about $10^{-1}$ were derived~\cite{VM}.

In the SM the Yukawa sector is both flavor-conserving and
CP-conserving, but these effects can be generated at the tree
level if new scalar fields are introduced. One alternative, which
does not contemplate the introduction of new degrees of freedom,
consists in incorporating into the classical action the virtual
effects of the heavy degrees by introducing $SU_L(2)\times
U_Y(1)$-invariant operators of dimension higher than four~\cite{EL2}.
Indeed, it is only necessary to extend the Yukawa
sector with dimension-six operators to induce the most general
coupling of the Higgs boson to quarks and leptons. A Yukawa sector
with these features has the following structure~\cite{EL2}
\begin{equation}
{\cal L}^Y_{eff}=-Y^q_{ij}(\bar{Q}_i\Phi_q
q_j)-\frac{\alpha^q_{ij}}{\Lambda^2}(\Phi^\dag \Phi)(\bar{Q}_i\Phi_q
q_j)+ H.c.,
\end{equation}
where $Y_{ij}$, $Q_i$, $\Phi_q$ ($\Phi_q=\Phi,\tilde{\Phi}$ for $q=d,u$ respectively and a sum over $q$ is implied), $d_i$, and $u_i$
stand for the usual components of the Yukawa matrix, the left-handed quark doublet, the
Higgs doublet, and the
right-handed quark singlets of down and up type, respectively.
The $\alpha_{ij}$ numbers are the components of a $3\times 3$
general matrix, which parametrize the details of the underlying
physics, whereas $\Lambda$ is the typical scale of these new
physics effects. This effective Yukawa sector induces a flavor and CP violating coupling $f_if_jH$ of renormalizable type given by $\Gamma_{f_if_jH}=-i(\omega^{ij}_RP_R+\omega^{ij}_LP_L)$, where $\omega^{ij}_R=\frac{gm_i}{2m_W}\delta_{ij}+\Omega_{ij}$ and
$\omega^{ij}_L=\frac{gm_i}{2m_W}\delta_{ij}+\Omega^*_{ij}$, with $P_{R,L}=(1\pm \gamma_5)/2$. In these expressions, $\Omega^{(u,d)}=\frac{1}{\sqrt{2}}\Big(\frac{v}{\Lambda}\Big)^2V^{(u,d)}_L\alpha^{(u,d)}V^{(u,d)\dag}_R$, being
$V^{(u,d)}_L$ and $V^{(u,d)}_R$ the usual unitary matrices, which correlate gauge states to mass
eigenstates. Let us emphasize that this vertex describes the most general renormalizable coupling of a scalar field to pairs of fermions, which reproduces the mean features of most of extended Yukawa sectors~\cite{AFRTTT}.

\begin{figure}
\centering
\includegraphics[width=3.0in]{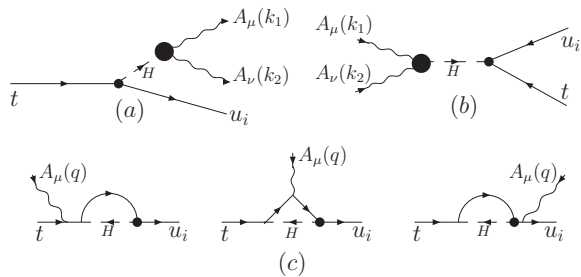}
\caption{\label{FD} Feynman diagrams contributing to the $t\to c
\gamma \gamma$ ($a$) and $t\to c\gamma$ ($c$) decays and to the
scattering process $\gamma \gamma \to tc$ ($b$).}
\end{figure}

\begin{figure}
\centering
\includegraphics[width=3.0in]{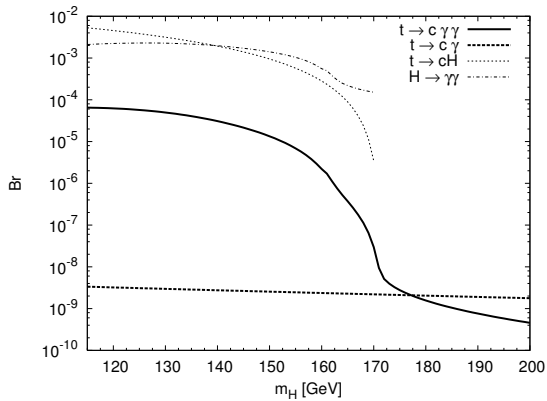}
\caption{\label{BR} The branching ratios for the $t\to cH$, $H\to \gamma \gamma$, $t\to c\gamma$, and $t\to c\gamma \gamma$ decays as a function of the Higgs mass. The thin lines correspond to a Higgs mass in the range  115 GeV $<m_H<$ 171.3 GeV.}
\end{figure}

The above $f_if_jH$ vertex induces the flavor violating vertices $f_if_j\gamma$ and $f_if_j\gamma \gamma$ at the one-loop level. The contribution to $f_if_j\gamma \gamma$ occurs through two set of Feynman diagrams, each of them giving a finite and gauge invariant contribution. One of these sets, whose contribution is marginal, include box diagrams, reducible diagrams characterized by the one-loop $f_if_j\gamma$ coupling, and reducible diagrams composed by the one-loop $f_i-f_j$ bilinear coupling (see Fig.~1 of Ref.~\cite{AFRTTT}). The other set of diagrams, which gives the dominant contribution, is characterized by the SM one-loop $\gamma \gamma H^*$ coupling, with $*$ stands for a virtual Higgs. The exact expression for the amplitude associated with the  $f_if_j\gamma \gamma$  vertex with all particles on the mass shell is given in Ref.~\cite{AFRTTT}. In the same reference, an exact expression for the decay width of the $f_i\to f_j\gamma$ transition is also presented. We will use these results to study the decays $t\to u_i\gamma$ and $t\to u_i\gamma \gamma$, as well as the scattering process $\gamma \gamma \to tu_i$ (see Fig.~\ref{FD}). As already mentioned, we are also interested in studying the $t\to u_iH$ decay, whose branching ratio can be written as $Br(t\to u_iH)=(\Omega^2_{tu_i}/(32\pi))(m_t/\Gamma_t)(1-(m_H/m_t)^2)^2$.
In writing the above expression, we have neglected the $m_{u_i}$ mass. The expression for the branching ratio associated with the $t\to c\gamma \gamma$ decay can be obtained from the exact expression given in Ref.~\cite{AFRTTT} for the $f_if_j\gamma \gamma$ coupling. We will consider only the contribution arising from the diagrams characterized by the one-loop $\gamma \gamma H^*$ coupling, as it dominates. In order to make predictions, some value for the $\Omega_{tu_i}$ parameter must be assumed. In Ref.~\cite{EL2}, the branching ratios for the $t\to u_iV$ decays were estimated by adopting the Cheng-Sher ansatz slightly modified by introducing the new physics scale $\Lambda$ instead of the Fermi one $v$: $\Omega_{tu_i}=\lambda_{tu_i}\sqrt{m_tm_{u_i}}/\Lambda$. By assuming $\lambda_{tc}\sim 1$, branching ratios for the $t\to cH$ and $t\to c\gamma$ transitions of the order of $10^{-4}$ and $10^{-8}$, respectively, were found. On the other hand, the branching ratio for the $t\to c\gamma \gamma$ decay was calculated within the context of the THDM-III, which has the simplest extended Higgs sector that naturally incorporates Higgs mediated flavor changing neutral current at the tree level~\cite{THDM-III}. Using the Cheng-Sher ansatz, it was found that this decays can reach a branching as high as $10^{-4}$. In this work, we will support our predictions using the available experimental data to get a bound on the $tu_iH$ vertex. We have found that the best constraint~\cite{FRRTT} arises from the recently observed $D^0-\overline{D^0}$ mixing~\cite{BaBar,Belle,HFAG}. Since this observable receives contributions simultaneously from both the $tuH$ and $tcH$ couplings, it is only possible to obtain a bound for the product of the two parameters characterizing these vertices, namely, $|\Omega_{tc}\Omega_{tu}|<10^{-3}$. So, in order to make predictions, some extra assumption must be introduced. We will assume that $\Omega_{tu}=10^{-1}\Omega_{tc}$, which implies the bounds $\Omega^2_{tc} <10^{-2}$ and $\Omega^2_{tu}<10^{-4}$. It should be mentioned that the experimental data on $D^0-\overline{D^0}$ mixing was used in Ref.~\cite{GHPP} to constrain this product of couplings within the context of the THDM-III. Although this reference presents a bound for the Cheng-Sher parameters ($\lambda_{ij}$ in our notation), we have verified that our bound for $|\Omega_{tc}\Omega_{tu}|$ coincides essentially with the one presented by these authors in the Higgs mass range $100$ GeV $<m_H<200$ GeV. However, we would like to stress that our parametrization of the $Hf_if_j$ coupling is more general than that generated by the THDM-III model in the sense that it includes both scalar and pseudoscalar components. Below, we will present results only for those processes involving the quark $c$, which can be translated to the corresponding transitions involving the quark $u$ simply multiplying by a factor of $10^{-2}$. We will present predictions using the maximum allowed value for the $\Omega_{tc}$ parameter. In Fig. \ref{BR}, the branching ratio for the $t\to cH$ decay and for the $H\to \gamma \gamma$ decay are shown as a function of the Higgs mass. From this figure, we can see that $Br(t\to cH)$ ranges from $5\times 10^{-3}$ to $5\times 10^{-4}$ for a Higgs mass in the intermediate range $115$ GeV $<m_H<2m_W$, which is of the same order of magnitude that the branching ratio for the $H\to \gamma \gamma$ decay. This prediction is about 1 order of magnitude above than that obtained in Ref.~\cite{EL2}. On the other hand, from this figure we can see that the branching ratio for the $t\to c\gamma$ decay is indeed suppressed, as it has a value essentially constant of about of $5\times 10^{-8}$, which however is about 6 orders of magnitude larger than the SM prediction. In contrast, the $t\to c\gamma \gamma$ decay can has a significant branching ratio in the intermediate Higgs mass range, which can be as high as $10^{-4}-10^{-5}$. This prediction is of the same order of magnitude as that obtained in Ref.~\cite{DPTT}. It should be noticed that $Br(t\to c\gamma \gamma)\sim Br(t\to cH)Br(H\to \gamma \gamma)$, which is evident from graphs in Fig.~\ref{BR}.

\begin{figure}
\centering
\includegraphics[width=3.4in]{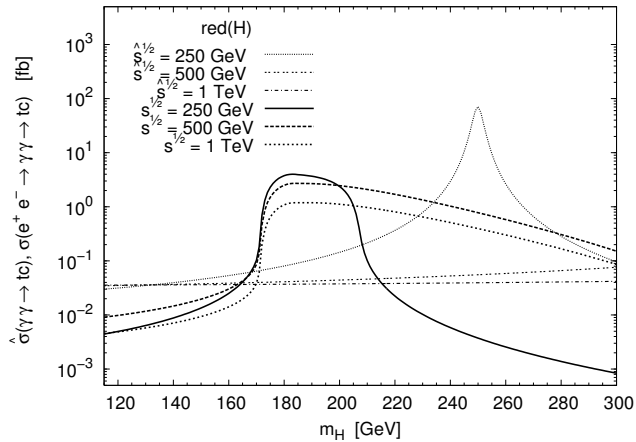}
\caption{\label{CSNE} The cross sections for the $\gamma \gamma \to t\bar{c}+\bar{t}c$ and $e^+e^-\to \gamma \gamma\to t\bar{c}+\bar{t}c$ processes as a function of the Higgs mass for some values of the c.m.s energy.}
\end{figure}

We now turn to discuss the $\gamma \gamma \to \bar{t}c+t\bar{c}$ process in the context of the ILC. From the amplitude for the $f_if_j\gamma \gamma$ coupling given in Ref.~\cite{AFRTTT}, one can construct the unpolarized cross section of the $\gamma \gamma \to \bar{t}c+t\bar{c}$ collision, which in turn defines the cross section of the complete process $e^+e^-\to \gamma \gamma \to \bar{t}c+t\bar{c}$ through the relation
\begin{equation}
\label{cs}
\sigma(s)=\int^{y_{max}}_{(m_{c}+m_t)/\sqrt{s}}dz\,\frac{d{\cal
L}_{\gamma \gamma}}{dz}\,\hat{\sigma}(\gamma \gamma \to \bar{t}\, c+t\, \bar{c}),
\end{equation}
where $z=\sqrt{\hat{s}}/\sqrt{s}$, being $\sqrt{s}$ and $\sqrt{\hat{s}}$ the c.m.s. energies of the $e^+e^-$  and $\gamma \gamma$ collisions, respectively. In the above expression $\frac{d{\cal L}_{\gamma \gamma}}{dz}$ represents the photon luminosity~\cite{Luminosity}. The optimum value for the upper limit in the above integral is $y_{max}\approx 0.83$. As in the case of the $t\to c\gamma \gamma$ decay, we are considering only the contribution given by the diagrams characterized by the SM $\gamma \gamma H^*$ vertex. Initially, the ILC will operate from a $\sqrt{s}$ of $200$-$500$ GeV, with a luminosity of $500\, \mathrm{fb}^{-1}$ within the first years of operation and $1000\, \mathrm{fb}^{-1}$ during the second phase of operation at $500$ GeV. After operating below or at $500$ GeV for a number of years, the ILC could be upgraded to a higher energy~\cite{ILC}. We will present results for energies in the range $250$ GeV $<\sqrt{s}<1000$ GeV. In Fig.~\ref{CSNE}, both the pure and the convoluted cross sections are shown as a function of the Higgs mass for some values of $\sqrt{\hat{s}}$ and $\sqrt{s}$. It can be appreciated a significant enhancement of the convoluted cross section for values of the Higgs mass near the value of the top mass. This peculiar effect, which, as it can be appreciated from the graphic in Fig.~\ref{CSNE}, is not present in the pure $\hat{\sigma}(\gamma \gamma \to t\bar{c}+\bar{t}c)$ cross section, has its origin in a resonance that occurs for a Higgs mass near the top mass value. To see this, note that the convoluted cross sections are proportional to the Higgs propagator $((sz^2-m^2_H)^2+m^2_H\Gamma^2_H)^{-1}$, which reach its maximum value for $z=m_H/\sqrt{s}$. On the other hand, from Eq.~(\ref{cs}) it can be seen that $(m_t+m_c)/\sqrt{s}<z<y_{max}$, which shows that the integration variable $z$ can be equal to $m_t/\sqrt{s}$ and thus a resonant effect arises when $m_H$ takes values near the top mass. The importance of the convoluted cross section for a Higgs mass of the order of the top mass is best illustrated by showing some specific values. For example, at a $\sqrt{s}$ of $500$ GeV the convoluted cross section takes the values $0.46\, \mathrm{fb}$, $2.72\, \mathrm{fb}$, $2.46\, \mathrm{fb}$, $2.09\, \mathrm{fb}$, and $1.70\, \mathrm{fb}$ for a Higgs mass equal to $m_t$, $184$ GeV, $200$ GeV, $210$ GeV, and $220$ GeV, respectively. The maximum value of the cross section corresponds to $m_H=184$ GeV. The respective number of events is 460, 2720, 2460, 2090, and 1700 if a luminosity of $1000\, \mathrm{fb}^{-1}$ is assumed. From the discussion given at the beginning of the paper, we can see that our results for $m_H$ near $m_t$ are about 1 order of magnitude larger than those derived from the THDM-III~\cite{THDM1,THDM2} and are lightly above of those predicted in supersymmetric models~\cite{SUSY1,SUSY2}. However, our results are 1 order of magnitude lower than those derived within the context of top-color-assisted technicolor models~\cite{TCAT} and several orders of magnitude lower than those obtained in the littlest Higgs model~\cite{LHM1,LHM2}. Our results are conservative since we have used the one-loop SM $H\gamma \gamma$ coupling.

In conclusion, in this paper we have investigated some phenomenological implications of a flavor violating $tcH$ vertex induced by an extended Yukawa sector that incorporates $SU_L(2)\times U_Y(1)$-invariant operators of up to dimension six. By bounding the $tcH$ vertex from the recently observed $D^0-\overline{D^0}$ mixing, the $t\to cH$, $t\to c\gamma$, $t\to c\gamma \gamma$, and $\gamma \gamma \to t\bar{c}+\bar{t}c$ processes were predicted. Sizable branching ratios for the  $t\to cH$ and $t\to c\gamma \gamma$ decays of the order of $10^{-3}$ and $10^{-4}$, respectively, were found. A branching ratio of about $5\times 10^{-8}$ for the $t\to c\gamma$ decays was found, which though it is about 6 orders of magnitude larger than the SM prediction, it is not large enough to be detected. As far as the $tc$ production at the ILC is concerned, it was found that about of $(0.5-2.7)\times 10^{3}$ $tc$ events may be produced for a Higgs boson with mass near to the top quark mass.

We acknowledge financial support from CONACYT and SNI (M\' exico).

\end{document}